# CityScopeAR: Urban Design and Crowdsourced Engagement Platform

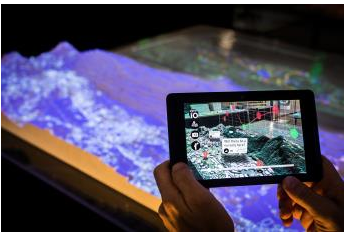

Figure 1. CityScopeAR along with a CS model of Andorra used to augment urban design proposals and facilitate pubic participation in the design process.

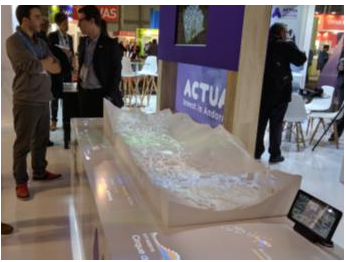

Figure 2. CityScopeAR on display at Smart Cities conference, Barcelona '16. The system recorded hundreds of daily user inputs and comments concerning design proposals.

**Ariel Noyman**
City Science Group
MIT Media Lab
75 Amherst Street
Cambridge, MA 02142 USA
noyman@mit.edu

**Yasushi Sakai**
MIT Media Lab
75 Amherst Street
Cambridge, MA 02142 USA
yasushis@mit.edu

**Kent Larson**
MIT Media Lab
75 Amherst Street
Cambridge, MA 02142 USA
kll@mit.edu

## Abstract
Processes of urban planning, urban design and architecture are inherently tangible, iterative and collaborative. Nevertheless, the majority of tools in these fields offer virtual environments and single user experience. This paper presents CityScopeAR: a computational-tangible mixed-reality platform designed for collaborative urban design processes. It portrays the evolution of the tool and presents an overview of the history and limitations of notable CAD and TUI platforms. As well, it depicts the development of a distributed networking system between TUIs and CityScopeAR, as a key in design collaboration. It shares the potential advantage of broad and decentralized community-engagement process using such tools. Finally, this paper demonstrates several real-world tests and deployments of CityScopeAR and proposes a path to future integration of AR/MR devices in urban design and public participation.

## Author Keywords
Urban Design; Mixed Reality; AR; Collaboration

## ACM Classification Keywords
H.5.m. Information interfaces and presentation (e.g., HCI): Miscellaneous; See http://acm.org/about/class/1998 for the full list of ACM classifiers. This section is required.

## Introduction: Limited Discussion

Urban design is the effort of many. Since the early days of Computational Aided Design (CAD), designers and engineers strived to create a seamless discourse between those who involved in city design. Despite technological advancements, processes of decision making for city design were left grossly by the hands of the few. Even in cases where public discussion was offered, limited collaborative tools were in the disposal of stakeholder and the general public. [sidebar (a)]

## Background: Compu-tangible Participation

Traditional CAD tools were commonly designed to offer working experience for individuals, with limited input (mouse, keyboard) and outputs devices (monitors, printers). Since the emergence of networking and shared working environments, tools have integrated collaboration capabilities such as sharing of documents or screens. Yet these capabilities were not fundamentally changing the design process; the act of sharing was following the act of design, but not in sync. [1] Even today, most tools still focus on a single user design environment. Additionally, it is rare that a tool is capable of engaging nonprofessionals, thus hindering the majority of potential users from having a say during a design decision process.

But starting from the 1960's, innovative design and technology pioneers strived to envision alternative design processes. The following key innovations questioned norms of collaborative design processes and set the context for the development of CityScopeAR.

E. Sutherland's revolutionary SketchPad (1964) became a cornerstone for CAD and HCI. Sutherland saw his tool as a potential mediator bridging between designers and engineers when collaboratively confronting design complexities. 2]

In 1970, MIT's Architecture Machine Group (AMG) combines physical and computational city-like models to an automated city design process called 'Seek' [4]. The project featured a computer-controlled environment with small, building blocks inhabited by gerbils. Following pre-programmed instructions, the robotic arm automatically rearranged and reacted to changes caused by the gerbils. AMG envisioned this as an 'architecture machine' that would turn the traditional human-machine dynamic into an iterative dialogue. This project highlighted that design and urban planning issues are socio-technical problems, and illustrated the importance of aiding systems capable of collaboration. [sidebar (b)]

Near the turn of the century, a series of Tangible-User-Interface projects tackled the complexity of real-time design and planning collaboration. The Augmented Urban Planning Workbench [3], The Clay Table [5], The I/O bulb and the Luminous Room [6] mixed traditional and computational design processes using tangible interfaces augmented by computational analysis. [sidebar (c)]

In 2004, The Bartlett has developed ARTHUR [7], an MR interface for urban design collaboration. Using Ericsson Saab Avionics glasses, ARTHUR displayed virtual models of a design scheme. A set of markers allowed users to physically interact and iteratively redesign the virtual model. ARTHUR team also proposed several urban simulations, so that design outcomes could be evaluated in real time. Aside from the novel usage of MR for urban design, ARTHUR favored collaboration and multi-user roundtable engagement as a key for successful design process.

---

*(a) "...The urban designer is in critical need of a platform that allows the simultaneous understanding of a wide variety of representations, including drawings, physical models, and digital analysis... simultaneous use of physical and digital media in the same space is an important requirement of the design studio of the future." [3]*

*(b) "The group is working on computer systems...not just to solve engineering problems, but to interact with the architect and discuss urban design problems with him." [4]*

*(c) "By 'triangulating' between these multiple forms of representation, we gain a more realistic sense of the site and proposed urban design... and the relationships between the static form of physical models and the dynamic behavior of previously intangible factors such wind speed, shadow movements and vehicular flow." [3]*

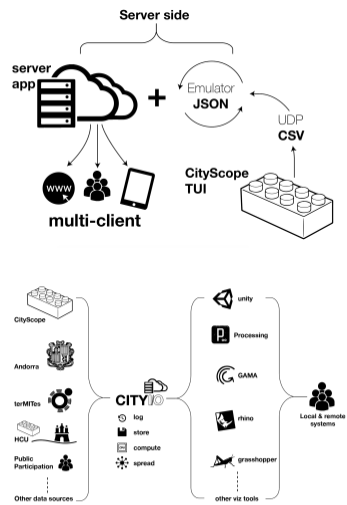

Figure 3. CityScopeAR, CS TUI and cityIO distributed architecture

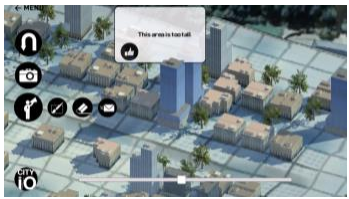

Figure 4. CityScopeAR UI during planning session: user observes real-time edits to the TUI, can scroll changes history back and forth and can add comments or 'likes' to comments made by others.

As shown, several advancements narrowed the gap between design tools and collaborative processes. Nevertheless, only few managed to weave together tactile design, real time computational analysis and community-wide feedback. In most cases, systems offered either broadly-shared but virtual environments (ARTHUR) or fixed position TUIs (Workbench). CityScopeAR offers a third option in which tangible interaction on discrete physical models is merged with location agnostic and scalable feedback system.

## Augmenting CityScope: System Design

Since '13, MIT City Science group is developing CityScope (CS), a TUI for urban modeling. CS platforms are designed for collaborative design processes though rapid prototyping and feedback loops. A generic CS setup includes an urban maquette, projectors and sensors to detect human interactions. CS has proven to assist decision making in diverse use cases involving urban design, transportation planning [1010], tourism analysis as well as refugees' accommodations [9]. Nevertheless, CS usability studies found several shortcomings: First, most CS platforms use flat-top tangible pieces projected with symbols and colors to portray various building types. When designers edit the layout, CS could not project three-dimensional building volumes or facades into thin air. In another case, users asked for different data layers to be to displayed simultaneously. But most importantly, CS had no mechanism to systematically collect, store and display users' feedback; at the time, traditional survey was tracking users' input during design sessions.
All these suggested that alongside the projection of base layers on the TUI, additional methods should be explored.

## CityScope with MR: Preliminary Prototype

CityScopeAR first iteration included a generic CS platfrom, laptop and webcam. The laptop displayed information layers that were mapped to the extents of the CS platform using colorful Lego as AR markers. These layers included enhancements to the urban model (such as 3d building facades, shade or vegetation) and data layers (urban analysis, heat maps, POIs). The tool was developed using Unity3d and vuforia SDK, later to be replaced by the open-sourced ARtoolkit and Nexus 9 tablet. This setup did not yet include a bidirectional communication between the device and the CS platform, thus limiting users to precomputed data and fixed set of visualizations.

## A Networked CityScope

Soon it became clear that in order to allow interaction and feedback, constant communication between CS and user devices was necessary. In mid '15, a preliminary communication stream between CS platforms and wireless devices was developed using the communication protocol UDP. Edits to the CS interactive elements were served to the device, triggering visual changes in MR environment.
UDP presented several challenges: data reliability was important than the speed, since lost packets caused inconsistent user experience. As well, UDP was not scalable to web apps, since it is blocked on modern browsers. Hence, a server-client system (called cityIO) has been proposed to allow scalable and reliable communication between elements in the CS ecosystem.

## Distributed Collaboration

CityIO was developed as a distributed server system, allowing multi-users to have bidirectional communication. Since '15, cityIO has been using standard HTTP, implemented in NodeJS. This has converted the CS ecosystem into a distributed platform that is extensible to multiple clients performing different tasks. This provided (1) an easy integration with powerful computation nodes, simulation tools or data queries and (2) reliable connectivity with different UI clients.

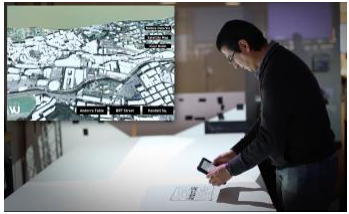
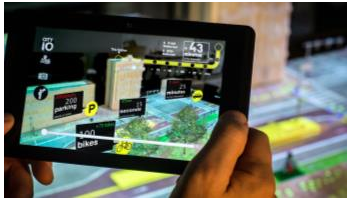
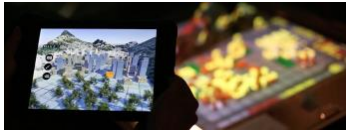
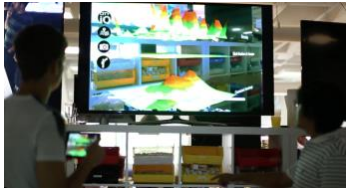

Figure 5. The evolution of CityScopeAR: (1) Static data display using AR markers (2) MR using object detection (3) Dynamic data (cityIO ver.1) with real-time updates (4) Tango device with screen casting and dynamic urban analysis using cityIO ver.2.

Utilizing this new bidirectional communication, end-users could now observe, interact and share their opinion during real-time CS design sessions. For example, users - both in proximity to CS platform or remote - could select geo-located spots in the virtual model, such as buildings, parks or roads and comment the designs proposed by other users. This type of distributed communication could offer designers, decision-makers and community leaders better understanding of the needs of the general public while avoiding cumbersome participation processes.

In early '16, CityScopeAR shifted to Google Tango Dev. Kit and SDK. This allowed better object recognition and preflight area-learning, especially effective in the challenging light condition imposed by CS projections. However, Tango's unique hardware and low market share contradicted the idea of decentralizing interaction by multiple users.

Since late '17, CityScopeAR is experimenting with Google's ARcore and Apple's ARkit. With built-in AR capabilities and without specialized hardware, these hold a promise for massive adaptation of AR by mainstream consumers. Furthermore, both frameworks are aiming to provide web browser ports which can lower adaptation barrier. If succeeded, tools such as CityScopeAR could eventually offer a native way to improve urban design collaboration and public participation.

**Tests and Deployments**

Throughout the development of CityScopeAR, several public deployments have offered the opportunity to test and gather feedback on usability and experience.

In early '15, City Science launched 'Boston BRT', A Bus Rapid Transit community planning process in the Boston metro area. Two CS TUIs were developed and used in community engagement events with over 300 participants. Following users' feedback, a CityScopeAR tools was designed to augment additional aspects onto the CS TUI: (1) BRT system performance information, such as trip duration, congestion or construction costs (2) an MR layer of 3d BRT station design, signage, building facades and vehicles.

In mid '15, a more advanced CityScopeAR was developed for the research corporation with the State of Andorra. Nicknamed 'AnodrrAR', this version was deployed at Barcelona's Smart Cities conference (Q1 '16) and later was set on permanent display at the Innovation Space in Caldea, Andorra (Q1 '17). AnodrrAR was designed as part of an interactive CS model of the city of Andorra La Vella. It depicted several data layers mapped onto the CS model: (1) Hi-res satellite imagery, 3d models of Andorra's landscape and current urban area (2) cell towers network simulation from which agent-based model was computed (3) design proposals for the downtown part of the city that were generated on the CS table. Using cityIO, user devices positioned and scaled changes to the tangible table onto geo-located anchors in the virtual city model (4) feedback system, allowing users to comment the urban development as well as other issues in the city.

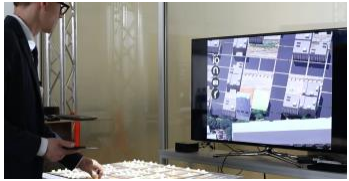

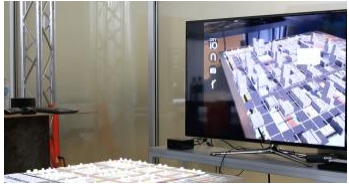

Figure 6. CityScopeAR deployment in HCU university, Hamburg. This version of the tool was built for the design of a district in the Rothenburgsort area. The user edits to the CS TUI are sent to a remote server. Then, an MR device translates these edits into a 3d model of the city and keystones it on top of the TUI extents. The resulting image is also casted to a large monitor so that other users can inspect the design. Urban parameters such as sun direction, building heights or land use could be adjusted on the both the TUI and user devices.

Since Barcelona's Smart Cities conference and the debut of Andorra's Innovation Space, hundreds of users were interacting with AnodrrAR. In Barcelona, nearly 200 comments were recorded locally during each day of the event. Similar iterations of CityScopeAR were deployed in Hamburg's HCU university ('16), Volpe redevelopment site in Cambridge ('17) and in Shanghai, Tongji Uni. ('17).

## Discussion and Contribution

The challenge of communicating complex design between professionals and nonprofessionals is in the core of decades of research. This paper presented the design, development and deployment of an MR/AR platform for urban design collaboration and public engagement. Its novelty is in the extension of a traditional design process, focusing on tangible interfaces and augmenting them on demand. As well, a distributed system for design discussion was introduced as a mean to widen the discourse.
Nevertheless, broadening the audience may raise deliberation costs exponentially. Despite current attempts to aggregate citizen virtual input into an inclusive design process, compu-tangible interfaces perform as solid evidence to complement integration.

## Conclusions and Future Work

The CityscopeAR Is a system that converges two worlds into one, tangible and MR interfaces to foster urban design processes. The tangible interface emits data through a web server to communicate with mobile devices. These can overlay additional information into an AR/MR environment as well as to allow bidirectional communication for design iterations or contributing comments and votes. We believe that further iteration using mainstream devices with Web Based AR frameworks will help lower the entry threshold and promote decartelize but tangible design experience.

## Acknowledgements
The authors would like to thank Nikita Samsonov (MIT EECS), Dalma Foldesi (MIT SA+P), Andorra team and the City Science group for their effort and support.